# CACHE REPLACEMENT ALGORITHM

**Submitted By**

**Sarwan Ali**
**(12PWBCS0303)**

**Supervisors**
**Dr. Iftikhar Ahmad**
**Siddique-ur-Rahman**

**August 2016**

**DEPARTMENT OF COMPUTER SCIENCE & INFORMATION TECHNOLOGY**
**UNIVERSITY OF ENGINEERING & TECHNOLOGY PESHAWAR**

# CACHE REPLACEMENT ALGORITHM

A thesis submitted in partial fulfillment
of the requirements for the award of degree of
BS(CS)
at
Department of CS & IT,
University of Engineering & Technology, Peshawar

by

Sarwan Ali
(12PWBCS0303)

In the supervision of
Dr. Iftikhar Ahmad
Mr. Siddique-ur-Rahman

August 2016

# Project Approval

This to certify that this project is approved and recommended as partial fulfillment for the award of Bachelor Degree in Computer Science, from University of Engineering & Technology Peshawar.

**Supervisor** ________________________________________

**Chairperson** ________________________________________

# DECLARATION

I, Sarwan Ali, announce that I am the sole author of this thesis. This is an authenticated copy of the thesis, including any required last modifications, as accepted by my supervisor. It is further proclaimed, that I have satisfied every one of the prerequisites in accordance with the Quality Assurance rules of the Higher Education Commission.

I understand that my thesis may be made electronically available to the public.

____________________

Sarwan Ali

12PWBCS0303

# Acknowledgements

As a matter of first importance, I am to a great degree thankful to Almighty Allah for making me prepared to do effectively fulfilling this research work. This postulation could truly not be finished without the kind help and colossal (extraordinarily) backing of various people. I might want to expand an additional ordinary expression of gratitude to my prestigious supervisor Dr Iftikhar Ahmad for his important time, most extreme inspiration and valuable remarks all through the advancement of my research. I am truly and powerful appreciative to him for helping me amid the review of the proposition, and for all the liberal remedies and corrections made by him. Subsequently it turned into a lighter, brief and doubtlessly succinct theory after his coordinated proposals. At last, a lot of appreciation is because of my family and companions, and particularly my folks to whom this theory is committed, for exceptional, unfaltering and superb backing all through my studies.

# Dedication

To my tender parents, relatives and my respectable teachers particularly Dr. Iftikhar Ahmad and Mr. Siddique-ur-Rahman, as they were supporting and spurring me all through my exploration. They all made me to stand here. Thanks a lot for the confidence in me, and for showing me so splendid standards of living.

# Abstract

Cache replacement algorithms are used to optimize the time taken by processor to process the information by storing the information needed by processor at that time and possibly in future so that if processor needs that information, it can be provided immediately. There are a number of techniques (LIFO, FIFO, LRU, MRU, Hybrid) used to organize information in such a way that processor remains busy almost all the time. But there are some limitations of every technique. We tried to overcome those limitations. We used Probabilistic Graphical Model (PGM), which gives conditional dependency between random variables using directed or un-directed graph. In our research, we exploited the Bayesian network technique to predict the future request by processor.

The main goal of the research was to increase the cache hit rate but not by increasing the size of cache and also reducing or maintaining the overhead. We achieved 7% more cache hits in best case scenario then those classical algorithms by using PGM technique. This proves the success of our technique as far as cache hits are concerned. Also pre-eviction proves to be better technique to get more cache hits. Combining both pre-eviction and pre-fetching using PGM gives us the results which were intended to achieve as the sole purpose of this research.

# Table of Contents

# Chapter 1 Introduction

There is a famous saying in computer science “There are only 2 hard problems in Computer Science, naming things and cache invalidation [1]”.

The cache memory is fast memory (25MHz to 5GHz speed) available inside the CPU with purpose to speed up the process of providing instructions stored in RAM memory directly to processor. It also helps to keep the processor busy for as maximum time as possible. By keeping the processor busy for the maximum time, the overall speed of the processor increases which increase the response time of the system. The cache memory is smaller in size, but faster which stores copies of the data required by the processor at that time or in near future from frequently used main memory locations.

Replacement policies in cache play significant part in determining the performance delivered by a cache and also by the processor. Previous researches on cache replacement policies observed a common behavior that a block of cache mostly becomes dead after some number of references by processor. This behavior is exploited by using different techniques. Each technique has its own points of interest and drawbacks.

## Background:

A computer is completely useless if we do not give instructions to the processor what to do. This is process done through a set of instructions, informing the processor that what next to do. Processor fetches needed data from random access memory (RAM). The main drawback of RAM is that when its power is cut, its data inside are lost, this property of RAM as “volatile”.

Therefore, instructions are needed to be stored on Hard disk, DVDs etc. (non-volatile), if they are required to be retrieved when PC is turned off.

When a program is executed, it is first loaded into the main memory, and then from the main memory, the processor loads that program through a circuit which is known as memory controller. It is located inside a processor.

The processor cannot fetch information from hard disk drive directly. The reason is slow speed of hard drives, even if we take fastest hard drive. For example, a SATA-300 hard disk drive has a maximum transfer speed of 300 MB/s. A processor with speed at 2 GHz along with 64-bit internal information-paths (the paths between the CPU internal circuits) will transfer information internally at 16 GB/s. This speed is over fifty times faster.

Processors have several different information-paths within the CPU, with every information path have different lengths, i.e. on AMD (Advanced Micro Devices, Inc) processors the information-path between the Level 2 cache and Level 1 cache is 128 bit. Similarly, on current Intel processors, this information-path is 256 bit. This variation in speed is because of the fact that hard drives are mechanical in nature, and are slower than electronics based technologies, because mechanical parts have to move physically to retrieve information. RAM is purely electronic. Therefore, it is faster than hard drives and optimally as fast as the processor.

Problem with the RAM is that the fastest RAM is not fast when it is compared with the processor’s clock cycle. Suppose we take DDR2-800 (DDR: Double Data Rate 2) RAM, it transfers information at the speed of 6400 MB/s to 12800 MB/s in case if the dual channel mode is current processors are capable of fetching information from the Level 2 cache at 128 bit (32 GB per second) or 256-bit rate (64 GB per second), only if the processor works internally at 2 GHz.

Transfer rates of information can be calculated using the formula given below (Here "data rate per clock cycle" is equal to "1"):

Transfer rate = width (number of bits) x clock rate x information per clock / 8

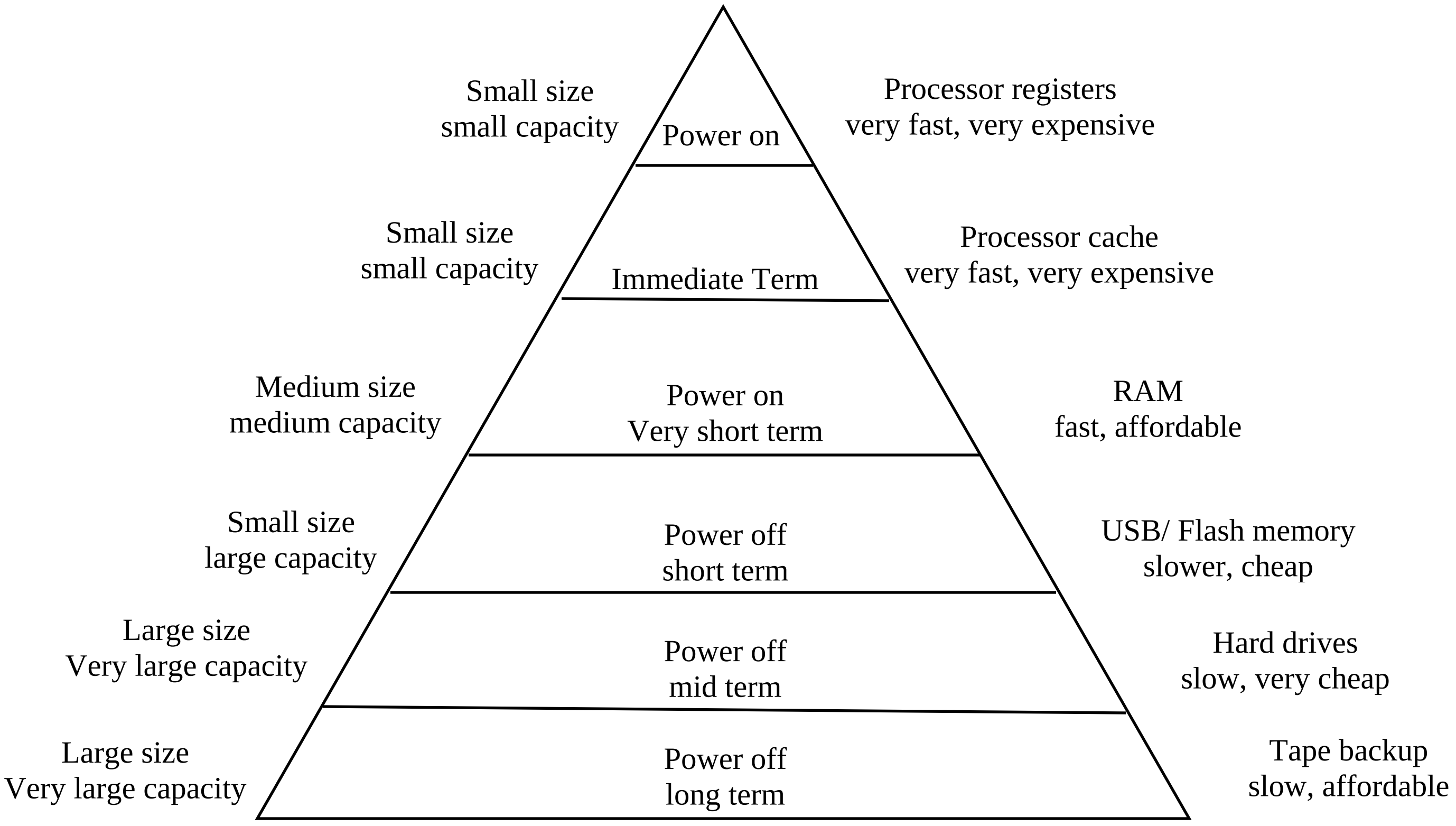

Figure 1.1 Memory hierarchy [2]

Figure 1.1 shows the different types of memories used inside computer system. The memories are closer to the processor in terms of their speed. The fastest memory is closest to the cache and so on. The reason is the higher speed of processor's clock cycle. Therefore, to match with the speed of processor, fastest memory is the most closet one. There are also drawbacks of every memory which are been shown in figure. The fastest memory which is register has speed same as CPU clock cycle. But it is small in size because of cost factor. Similarly, as we go away from processor towards slow memories, the cost start reducing therefore size starts increasing but on the other hand speed also reduces.

## 1.1. Problem:

The problem of cache less system is not only latency (latency is the amount of time a message takes to traverse a system), but also the transfer rate. Latency (“access time”) can be defined as memory delaying time to give the information back to the processor which was requested, this is not instantaneous. When the processor requests as information which is stored at specific address, a certain time is comes from memory to give this information back. If memory is labeled as having a CL (Column Access Strobe Latency) of 5, then it means that the required requested information will be delivered memory only after 5 memory clock cycles.

Wait decreases the processor performance. In the event that the CPU needs to hold up five memory clock cycles to get the direction or information it solicited, its execution will be just 1/5 of the execution it would get in the event that it was utilizing a memory equipped for conveying information promptly. As it were, while getting to a DDR2-800 memory with CL5, the execution the processor gets is the same as a memory working at 160 MHz (800 MHz/5). In this present reality the execution abatement is not that much since recollections work under a mode called burst mode where from the second information on, information can be conveyed promptly, on the off chance that it is put away on a touching location (as a rule the directions of a given system are put away in consecutive addresses). This is communicated as "x-1-1-1" (e.g., "5-1-1-1" for the memory in our case), implying that the principal information is conveyed after five clock cycles yet from the second information on information can be conveyed in only one clock cycle – in the event that it is put away on an adjacent location [3].

Cache Memory was first used in computer system at the 386DX timeframe. There are three types of Cache typically L1 (level one cache), L2 (Level two Cache), and L3 (level three cache) cache. The speed for these cache memories is 1ns, 4ns, and 25 ns.

## 1.2. Objectives:

The problem with the cache memory is that there are fewer cache lines than memory blocks so our objective for the thesis is to design an

- Efficient algorithm for mapping memory into cache lines
- A mean to determine which memory block is in which cache line.

There are two problems which are consider important when we talk about cache algorithm

1. Cache Miss
2. Cache Hit

### 1.2.1. Cache Miss:

When processor needs a specific piece of information, it first requests to the cache memory, the information is then searched in multiple cache levels i.e. L1, L2, L3, if the information is not found in cache, then it is considered as a cache miss. Then the information is fetched from the main memory and firstly it is stored in cache then it is given to processor.

### 1.2.2. Cache Hit:

A cache hit is a situation where the information solicit for processing by a part or application is found in the cache. We can say that it is a faster means of delivering information to the processor, because the cache memory already contains the solicited information.

## 1.3. Goal

The purpose of this research is to establish a new method for cache miss problem based on probabilistic graphical model to design an algorithm which can prognosticate future items that has highest probability to be called in near future. Further it was intended to study other techniques to get an overview of their pros and cons so that those mistakes can be avoided in our research. Based on sampling techniques, joint usage of statistical techniques and the knowledge of cache analysis is utilized to develop this technique. Furthermore, this method will not only permit a prognostication of the *mean miss rate,* but will also provide an empirical estimate of the actual distribution of the miss rate.

## 1.4. Thesis Organization:

Chapter 2 consist literature review in which we analyze different existing solutions and their limitations. In Chapter 3, we present our proposed approach which we used to solve the existing cache problem. In Chapter 4, we discuss the experimental setup and provide discussion on the results. In Chapter 5, we conclude the work and provide directions for future research.

# Chapter 2: Prelude and Literature Review

## 2.1. Cache Memory Problem:

Cache memories can add to huge execution favorable circumstances because of the gap between the processor and speed of memory. They customarily are considered as supporters to capriciousness on the grounds that the user cannot make sure of precisely how much time will be consumed when a memory operation is executed. In an ongoing framework, the cache may add to a missed deadline by really slowing the systems speed, however this is uncommon. To stay away from this issue, the engineers of run time systems have run the instructions in the way it was done before; with disabled cache, as a sanity check. Turning off the cache, nonetheless, will likewise make different components for example instruction pipelining less gainful, so the new system's execution speed will not be as good as it was intended to give

The method to determine the boundaries in computer in terms of execution time with cache memory were represented first in 1980's that is 20 years after the first cache memory was designed. Now after 15 years, there are a lot more methods which are more reliable and fast are developed in order to determine execution time with cache [4].

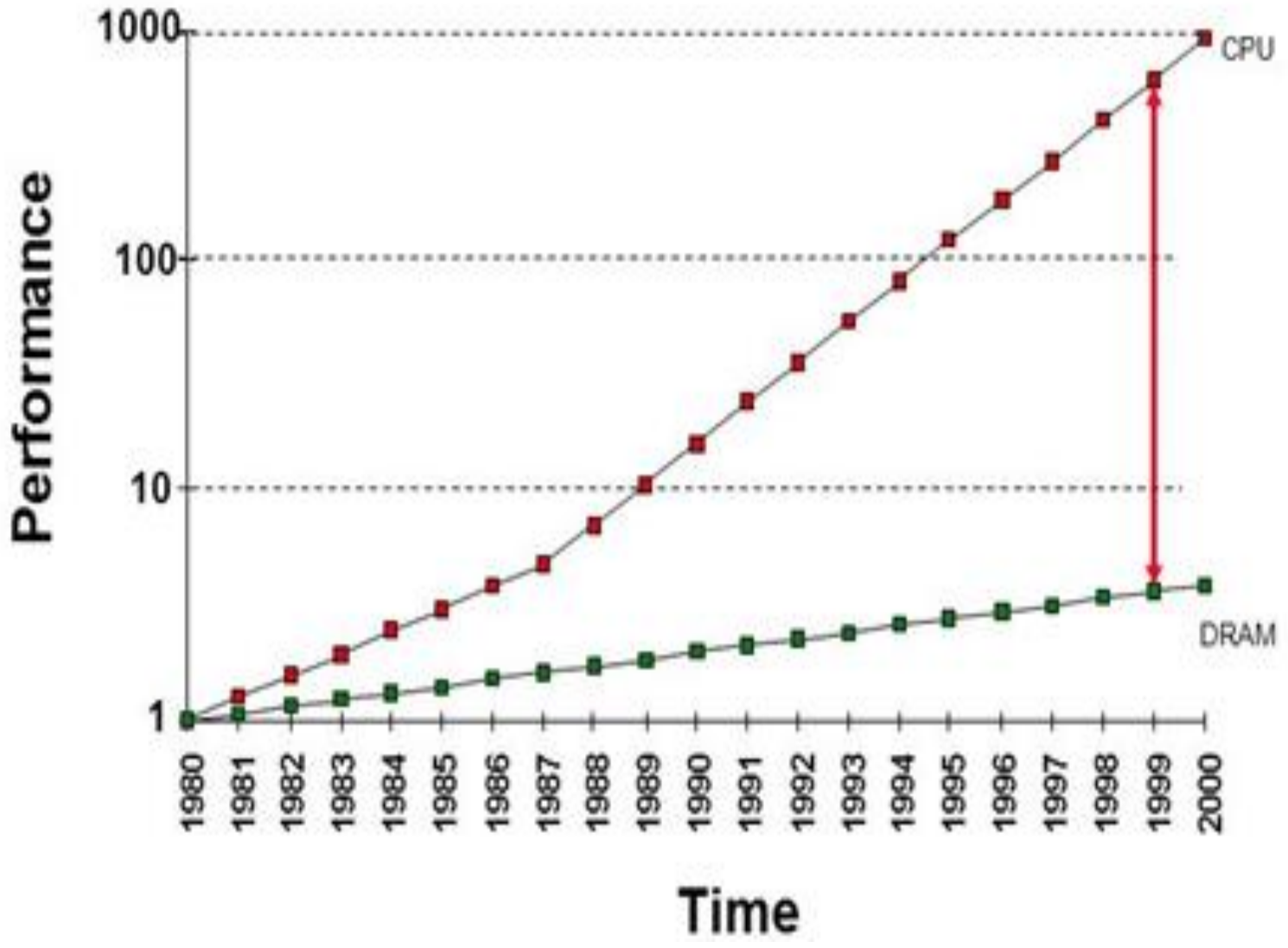

Figure 2.1 Processor memory gap [5]

In Figure 2.1, we can see the improvement in performance of processor starting from 1980 to 2000. The processor and memory gap performance is increased with the rate of 50% per year. The speed of DRAM is increased with the passage of time but not with that much rapid speed as compare to the speed of processor. Because of the high speed gap between RAM and processor, there was a need to introduce a type of memory which can be comparable with processor in terms of speed. Therefore, cache memory was introduced. After the introduction of cache memory, there were problems with cache memory, main of which is its cost. Cache memory capacity cannot be increased because of its higher cost. Therefore, an efficient algorithm was needed in order to use small size cache as efficient as possible.

## 2.2. LRU Algorithm:

LRU (Least Recently used) is a group of caching technique, which disposes of the slightest as of late utilized item first. This algorithm requires monitoring when the thing was utilized, which is costly in the event that one needs to ensure the algorithms dependably disposes of the slightest as of late utilized thing. General usage of given method require that age bits will be kept for cache lines and Slightest Recently Used cache line will be tracked taking into account age-bits.

The key data structure of the algorithm is the mix combination of **Doubly Linked List** and **Hash Map**. Doubly Linked List has been utilized in order to index the pairs in the order of information age, and initialize a Hash Map in an already defined length to store pairs.

When the information with Key E is queried, the function get E is initially called. On the off chance that the information of E is in the cache, the information is just returned then by cache and invigorate (Refresh) the information in the linked list. In order to invigorate information I in the list, we move the item of information I at head. If that is not the case then, if the information I of key E is not in the cache, then pair is needed to embed into the list. In the event that the cache is not full, it is embedded into the hash map, and the item is added to the head. On the off chance that the cache is as of now fully consumed, we get the tail of the list and update it with, then this item is moved to the head of the list [6].

## 2.2.1. LRU Example:

Reference String: 1 2 3 4 1 2 5 1 2 3 4 5

| | | | | |
|---|---|---|---|---|
| 1 | 1 | 1 | 1 | **5** |
| 2 | 2 | 2 | 2 | 2 |
| 3 | **5** | 5 | **4** | 4 |
| 4 | 4 | **3** | 3 | 3 |

In given LRU example, we can see the working of least recently use technique by giving a reference string to the algorithm. The item which has been used for the minimum number of time is replaced with the new item in the reference string. The replaced items has been shown in the figure.

### 2.2.2. Input:

The input will be a progression of information sets, 1 for every line. Every information set contains a string of multiple characters and an integer *N*. That integer represents the size of the cache. Consecutive characters comprise exclusively of capitalized letters and shout marks (exclamation marks). An access to that specific information is represented by a capitalized. A shout mark represents a solicitation to print the current values of cache.

The info must be progression of information sets, single for every line. Every information set will comprise of a whole number and a string of multiple characters. The whole number speaks to span of store for information set. Series of characters comprises exclusively of shout marks and capitalized letters. The letter speaks to an entrance to that specific bit of information. A shout mark speaks to a solicitation to display the present substance of the reserve.

Taking an example, the succession GHI!JKGL!H! Intends to get G, H, and I (in that order), print the substance of the cache, access J, K, G, and L (in specific order), then print the substance of the cache, then get to H, and again print the substance of the cache.

The order will every time starts with a capitalized letter and contain at least one shout mark.

If line contain number “zero”, then end of information will be signaled

### 2.2.3. Output:

For every information set output ought to be the line "Simulation S". Value of S is

1. One for the first data set

2. Two for the second data set, and so on.

At that point for every shout mark in the information set output ought to be the contents of the cache on single line as order of characters which will be representing multiple information set at present in the cache. Characters ought to be sorted all together from least used items to most used items, with slightest as of late happening first. Output ought to only be the letters which are present in cache; along with the condition that the cache is not full, in that condition there basically will be less characters to give the output (empty spaces will not be printed). The important thing to note is that in light of the fact that the order always starts with a capitalized letter, system will never request to display an output with completely empty cache [7].

| Input | Output |
|---|---|
| 5 GHI!JKGL!H! | Simulation 1 |
| 3 OPOQR!QROQP!PQPQ! | GHI |
| 5 KMKMN! | IJKGL |
| 0 | JKGLH |
| | Simulation 2 |
| | OQR |
| | OQP |
| | OPQ |
| | Simulation 3 |
| | KMN |

## 2.3. FIFO algorithm

The low-overhead and least difficult replacement algorithm is the FIFO algorithm. FIFO is otherwise called as a round robin. The way it works is when the cache is full it simply replaces the first item that was placed in the cache with the item needed by processor at that time, and the next replacement then will be the 2nd item placed in the cache and vice versa.

There is no single cache algorithm which will dependably perform well since that requires impeccable learning of the future. The strength of LRU in VM cache design is the aftereffect of a long history of measuring system behavior. Given genuine workloads, LRU works quite well a substantial division of the time. Be that as it may, it is not hard to build a reference string for which FIFO would have better execution performance over LRU.

We can create a FIFO queue to hold every item present in the main memory. At the head of the queue we supplant (Replace) the item. We embed item at the tail of the queue when an item is added in into the memory disk.

### 2.3.1. FIFO Example:

Let us consider the following string:

7 0 1 2 0 3 0 4 2 3 0 3 2 1 2 0 1 7 0 1

Buffer size: 3

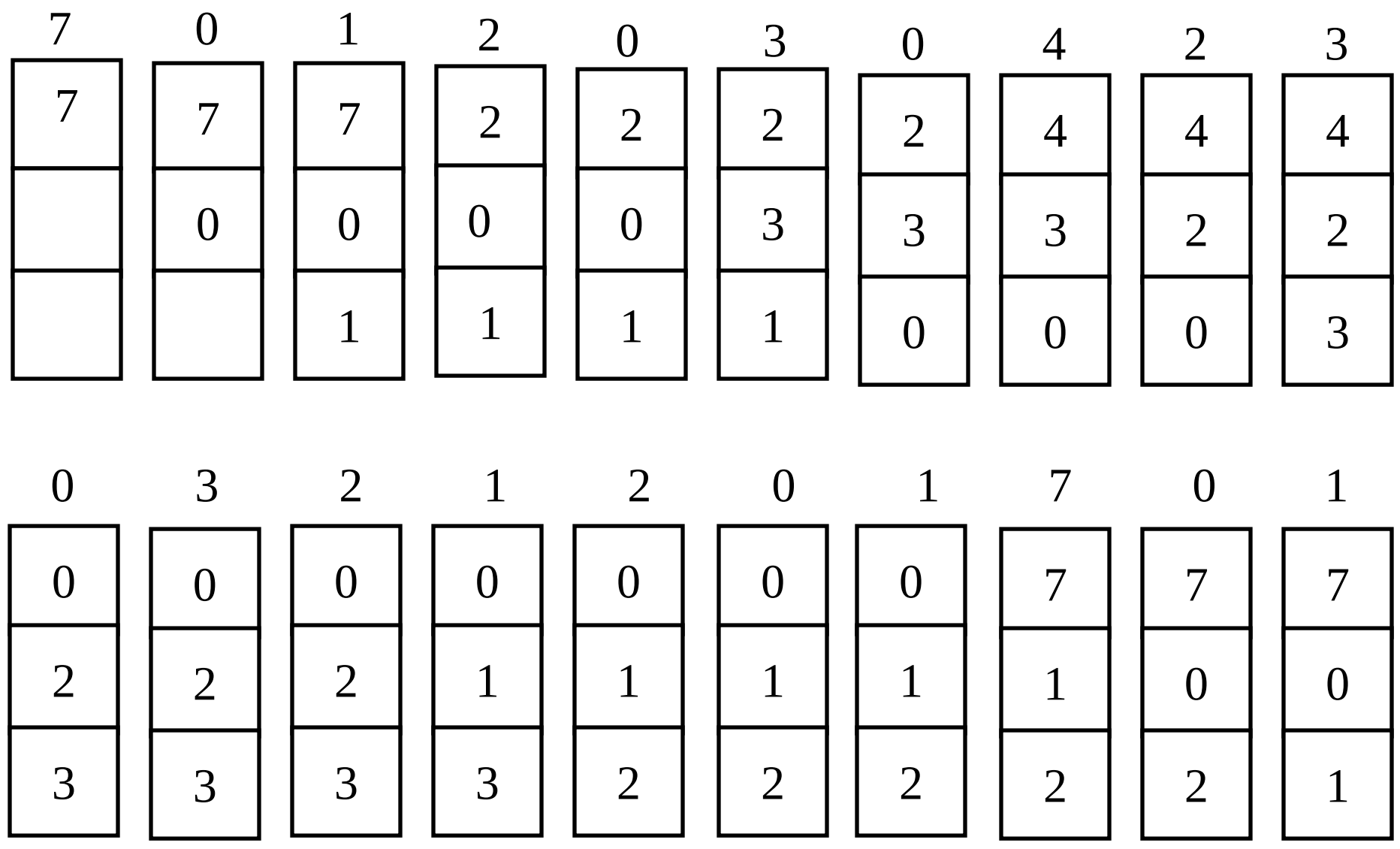

Total item fault occurs: 15

At start the three frames are empty. The first three references (7, 0, 1) causes cache miss and are added the items into the three empty places. The next item is “2” which supplant the item “7” because that was added first. Since “0” is the next item in the string but “0” is already in the memory. Therefore, there will be no cache miss for this item. This process goes like this as shown in the example. For every instance a cache miss occurs, it has been shown that which item are in all three frames at any specific time. There are overall 15 cache miss at the end.

## 2.4. LIFO Algorithm

A real-world stack permits operations toward one side only. For instance, we can place or expel a card or plate from top of the stack as it were. Moreover, Stack ADT (Abstract data type) permits all information operations toward one side as it were. At any given time, we can just get to the top component of a stack. This feature makes it LIFO data structure. The element which is placed (embedded or added) last, is accessed first. In stack terminology, embedded operation is called PUSH and evacuation operation is called POP.

LIFO works just opposite to that of FIFO. The item that comes last in the cache memory is eliminated first from it, regardless of the fact that it can be needed by processor in near future. In this way, cache miss rate can be increased because items are evicted from the cache not on the basis on their priority. The working of LIFO is shown in Figure 2.4.

## 2.5. Pseudo-LIFO Algorithm

Pseudo-last-in-First-out, is essentially new technique of replacement policies that controls very cache set as single fill stack (instead of conventional traditional access recency stack). There are three techniques used in this algorithm,

- Dead block prediction LIFO
- Probabilistic escape LIFO
- Probabilistic counter LIFO

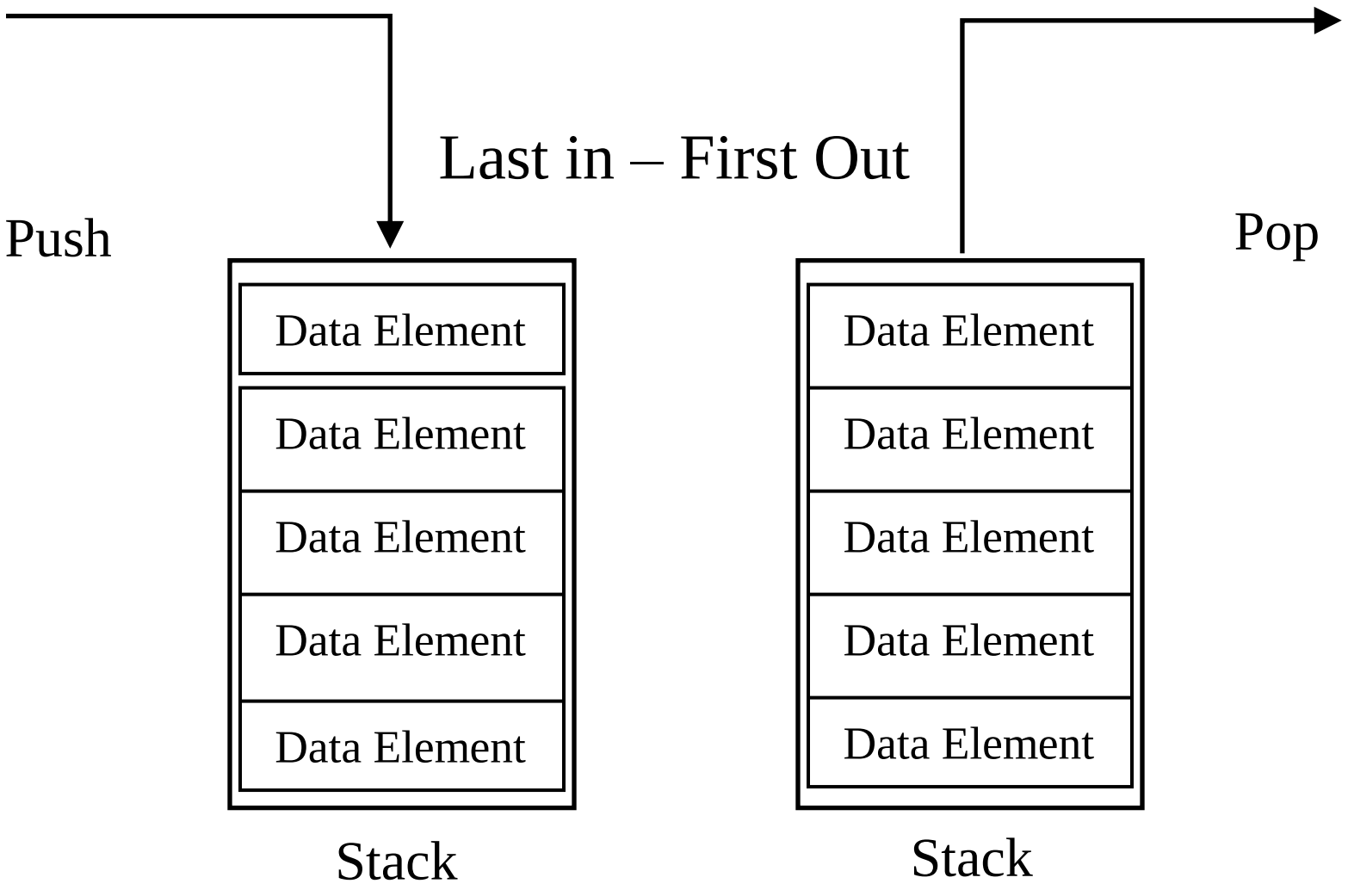

Figure 2.4 Working of stack (LIFO approach) [8]

### 2.5.1. The Dead block prediction LIFO:

dbp-LIFO (Dead block prediction LIFO) approach utilizes a dead block prognosticator (predictor). This methodology keeps up nineteen bits of additional situation per cache block when we compare it with the pattern: 6 bits of reference check , the lower 8 bits of the system (in wake of expelling the least significant 2 bits), 4 bits for fill stack position, and a dead bit. Every 1 MB Level 2 cache memory is equipped with single 2048-arrival history table for every core, for the purpose to carry out dead block prognostication. This table is listed with 8bits of system chain after 3 lower bits of the block address (in wake of expelling the bank id bits). Every table section contains single legitimate bit,

1. Single hysteresis bit
2. 6 bits of reference count
3. 6 bits of filter reference count

The conventional dead block prognosticator based replacement algorithm needs 37 KB for single-threaded, 232 KB for multi-programmed, and 172 KB for multithreaded simulation environment of auxiliary storage. For dbp-LIFO, these prerequisites are 198 KB for multi-programmed, 264 KB for single-threaded, and 45 KB for multi-programmed.

### 2.5.2. Probabilistic escape LIFO:

Pe-LIFO (Probabilistic escape LIFO) powerfully takes in the utilization probabilities of cache memory exceeding every available fill stack position with purpose of executing another substitution methodology.

## Pe-LIFO Properties:

1. Storage overhead is 1% of pe-LIFO, which diminishes the execution time by 10% on average when we compare it to a LRU methodology for single set contains 14 single threaded but not multithreaded applications on a two MB sixteen way set acquainted Level 2 cache.
2. The normal CPI is lessen by 19% by pe-LIFO on all things considered for an arrangement of 12 multi-programmed workloads in the mean while fulfilling single solid reasonableness necessity on a multiprocessor containing 4 core chip with an eight MB sixteen way set cooperative common Level 2 cache memory.
3. It diminishes by 17% the parallel execution time on all things considered for an arrangement of 6 multithreaded programs which is based on multiprocessor containing 8 core chip with single four MB sixteen way set acquainted common Level 2 cache memory.
4. PeLIFO methodology has storage overhead of one- 5th to half as compared with cutting-edge dead block prognostication-based replacement methodology. In any case, the pe-LIFO methodology conveys improved aggregated performance for the chosen multi programmed and single threaded workloads and comparatively aggregated pursuance for the multithreaded workloads contrasted with the replacement methodology based on dead block prognostication [9].

## 2.6. MRU Algorithm

MRU works totally inverse to that of LRU. The items which has been used most number of times is removed from cache when cache memory is full and processor solicit for an item that is not in cache. This approach has a major drawback that the item which has been called most number of time has the highest probability to be called again. Therefore, that item must remain in the cache for the longest timeframe. But despite of being inside the cache, those items are removed which results in the high cache miss rate.

MRU has the issue that in the event that we get a hit on a section that is frequently utilized as a part of "normal" mode while doing MRU lookups, we wind up tossing out the entry. Better other options to MRU for doing a scan without contaminating the cache are following:

- Bypass/Sidestep the cache totally,
- Test the cache without doing a read through/overhaul, and without changing the LRU chains.

## 2.7. Hybrid approach

ARC is a hybrid cache replacement algorithm with comparatively effective execution than that of LRU [10]. This is proficient by monitoring of both frequently used and recently used items addition to a recent eviction/removal history for both.

ARC keeps up 2 LRU items lists:

1. L1 (It maintains those items which are seen only once at a recent time)

2. L2 (It maintains those items which are seen at least twice, at a recent time)

The set of instructions really caches just a small amount of the items on these available lists. The items which appear twice inside brief span may have considered as of having recurrence or distant future multiple use potential. Subsequently, therefore it can be said that Level 1 catches recency, and Level 2 catches recurrence /frequency.

On the off chance that the cache can hold c items, these 2 records has been attempted to be kept generally of the same size by us, which is c. jointly, the 2 records involve a cache index that holds maximum of two c items. ARC technique caches a variable number containing latest items from Level 1 and Level 2 so that the aggregate number of items which has been cached is c. This technique consistently adjusts exact number of items from every cached list.

To balance an adaptive methodology with a non-adaptive methodology, assume $FRC_P$ (Fixed replacement cache) gives a settled substitution approach that endeavors p latest items from Level 1 to keep them in cache and also c−p latest items in Level 2. Subsequently, ARC carries on similar to $FRC_P$ with the exception that it can change value of p adaptively. A learning tenet/rule presented by us that gives ARC a chance to adjust viably and rapidly to a single changing workload.

Numerous programming techniques use recurrence and recency as prognosticators of the probability that items will be used again later on. ARC technique also behave as adaptive filter to track and recognize locality of reference (i.e. temporal locality). On the off chance that either recency or recurrence turns out to be more essential sooner or later, ARC technique will track the change which has been take placed during all process and adjust its interest both lists in like manner.

ARC perform its operations similar to that of FRC$_P$ methodology, notwithstanding when FRC$_P$ methodology utilizes hindsight to pick leading settled p regarding any specific cache size as well as the workload. Shockingly, ARC algorithm that works totally online, conveys execution performance which we can say that is as good as few cutting-edge cache replacement algorithms, notwithstanding when, with knowledge of the past, these techniques pick the best available fixed values to use for tuning parameters. ARC matches LRU's effortlessness of utilization, requiring only 2 LRU records.

The ARC research was carried out in research paper "Adaptive Insertion Policies for High Performance Caching" [11]. ARC was recently developed in light of the destructions of LRU approaches. LRU catch just recency but not recurrence/ frequency, also it further can be effortlessly "dirtied" by a scan. We can define scan as an order which can be utilized just once the item solicitations, and prompts to less and degraded performance. ARC beat defeats these two ruins by utilizing 4 doubly linked lists. The linklists L1 and L2 are what which are really in the cache memory giving any specific duration. On the other hand D1 and D2 go about as a second level. D1 and D2 have items that which are tossed out from L1 and L2 separately.

The aggregate number of items in this way anticipated that would actualize these rundowns is 2 x Q, where Q is the quantity of items present inside cache. L2 and D2 have just items which are utilized more than single time. These lists uses both LRU substitution, which have items who are expelled from L2, and those are then placed in D2. L1 and D1 work likewise side by side, with the exception of time which contain a hit in L1 or D1 then item is transferred to L2. The reason because of which this methodology is really extremely versatile is the total measures of the lists modified. The size of List L1 and L2 scrutinize dependably added into the aggregate number of items inside cache. Regardless, suppose that a hit occurs in D1, which is also called Phantom Hit. It expands the measure of L1 by 1 and lessening the measure of L2 by 1. In the other bearing, a hit in D2 “Phantom Hit”. This surely increase the measure of L2 by 1 and lessening the measure of L1 by 1. This permits the cache to adjust in accordance with having either more recency or more recurrence depending upon the workload [12].

Following accompanying request demonstrates how ARC out plays out the majority of alternate policies.

e.g. 0 1 2 3 0 4 1 2 3 0 1 2 3 0 1 2 4 3

Number of items = 3

The order beneath shows that ARC does not generally have the best execution however in more than one cases the quantity of cold start misses is lessened.

e.g. 1 2 5 4 3 2 1 5 4 6 8 4 5 2 7 10 1 2 15 15 2 4 6 3 2 21 8 78 6 1 2 3 4 5 6 7 8 9 1 3 22 5 4 9 15 12 9

Number of items = 10

### 2.8. PGM approach:

PGM (Probabilistic graphical model), otherwise called belief networks or Bayesian networks, are as of now settled as representations of domains including unverifiable relations among few irregular/random variables. To some degree less very much established, however maybe of equivalent significance, are dynamic probabilistic networks (DPNs), which demonstrate the stochastic development of a set of random variables after some duration. DPNs have huge points of interest over contending presentations i.e. Kalman filters, which look after concealed Markov model along with uni-modal posterior conveyances and linear models, which contain parameterization which grow aggressively along the quantity of state variables [13].

The common techniques used in PGM are

1. Markov chain
2. Bayesian Network

We used this approach in designing our algorithm in order to check its behavior and compare them with traditional approaches. PGM is used to show the relationship between random variables by showing their conditional dependencies.

In PGM technique, we analyzed its techniques, methods, approaches and different ways in which we can use it in our research. After evaluating this approaches, we designed our approach and experimental setup so that we can achieve our goals.

In Figure 2.8, we can see the conditional dependencies shown using PGM technique between three different identities which are closely related to each other. Their results using all possible combinations has been shown in Figure 2.8, using which we can easily find out that which identity has the highest probability to occur next in a certain scenario.

Sprinkle

| Rain | T | F |
|---|---|---|
| F | 0.4 | 0.6 |
| T | 0.01 | 0.99 |

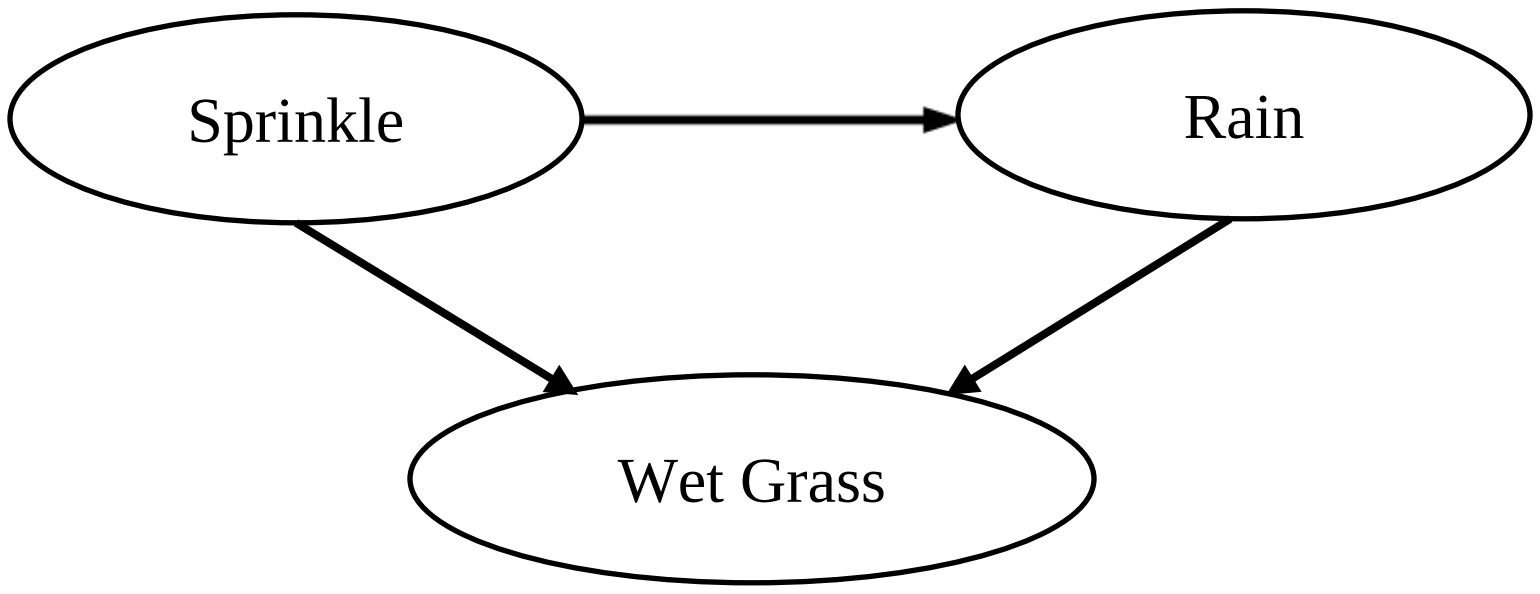

Rain

| T | F |
|---|---|
| 0.2 | 0.8 |

Wet Grass

| Sprinkler | Rain | T | F |
|---|---|---|---|
| F | F | 0.0 | 1.0 |
| F | T | 0.8 | 0.2 |
| T | F | 0.9 | 0.1 |
| T | T | 0.99 | 0.01 |

Figure 2.8 PGM example [14]

# Chapter 3 Proposed Approach

We used customized approach to make our research and experimental setup.

In our customized model, we used two techniques.

- Pre-Eviction
- Pre-Fetching

## 3.1. Pre Eviction:

An Eviction methodology is a class that precisely has knowledge that how to handle eviction occasions to track the action in its district. It might have a specific arrangement of setup properties that give it rules for when a specific node in the region ought to be ousted/removed. It can then utilize that configuration and also knowledge of activity it has in the area to figure out what nodes to expel/remove.

We propose cache replacement technique using Pre-eviction in which we evict elements using **locality of reference** [15]. Using temporal locality, we eliminate the items which are in the cache more than the selected CPU cycles and still is unused.

### 3.1.1. Method of Using Locality of reference:

1: Firstly, we suppose that locations of items in the memory for an application is in specific order (ascending or descending), in that case we will use "Halfway" approach. If an item in the cache has location greater than the half of the total item for that application in memory, then all the items above that specific item location will be evicted.

The reason is that if a specific memory area/location is referenced at a specific time, then it is likely that adjacent memory area will be referenced in a not so distant future. And suppose that items are arranged in ascending order, then it is possible that all items above specific items will not be called in near future (Instead below items will be called probably). Therefore they all are evicted.

2: Further we will also use another type of locality which is temporal locality, keeping a constant time, if an item is failed to get a hit in that specified time, then that item is evicted. Also it means that other items related to that evicted items has very less chance to be called. Therefore, they can also be evicted.

### 3.1.2. Null Eviction Policy:

The Null-Eviction-Policy is a basic methodology that effectively do nothing to perform any operation. It has been designed to be used to proficiently short-circuit eviction handling for areas in which we do not need objects to be removed (As an example, the timestamps cache, which ought to be never contain the information which is removed/thrown out). Just because the Null-Eviction-Policy does not really remove anything, therefore it does not take any type of design parameters [16].

### 3.1.3. Pre eviction Algorithm:

The algorithm to implement pre-eviction policy is as following.

```
Initialize 2 queues Block1, and Block2;
Both queues are initially empty;
If (item[i] in the cache) {
Return;
}
Else if (item[i] does not fit in queues tail)
{
```

```
If (there is no free block && item[i] location is greater than halfway then)
     {
          Assign Block1 values < halfway;
          Assign Block2 values > halfway
          Block1=null;
     }
If (Time of item[i] =0)
     {
          Remove item[i];
     }
}
```

If item is found in cache, then it is returned. If item is not in cache, then this because of one of the two factors

1. Cache is empty
2. Cache is full and required item is not present

This will cause a cache miss. In these condition, we will use our logic.

### I) First Approach:

In first technique, we used half way approach. If the called item called by processor has location greater than or equal to half of the total size of queue, then items are divided into two queues and items in first queue are evicted. The reason is that according to locality of reference, those evicted items have least probability to be called again.

### II) Second Approach:

In second approach, we used temporal locality. In this approach, if an item is unable to get a cache hit and its timer becomes zero (i.e. for 2K requests), then that item is evicted.

## 3.2. Pre Fetching:

In computer science, pre-fetch is a procedure utilized as a part of CPU to accelerate the execution of an instructions by lessening holding up time. Prefetching happens when processor demands an instruction or data block from main memory before it is really required. Once the block comes back from memory, it is set in a cache. At the point when an instruction is required really, it can be referenced inside a great deal more less time from the cache than if it needed to make a solicitation (Request) from RAM (random access memory). In this way, prefetching shrouds memory access latency (The amount of time a message takes to traverse a system.) and henceforth, it is a helpful strategy for addressing the memory wall (It can be described as a growing inconsistency of speed between processor and other memory which are outside the processor chip) issue. Pre-fetching is done in both of two circumstances

1. Demand misses: Fill solicitation due to cache miss
2. Pre-fetch: Fill solicitation in anticipation of data solicitation

Pre-fetching is of great importance because even if cache miss is not totally avoided, cache miss latency is reduced. Pre-fetch coverage is the percentage of misses avoided due to pre-fetching. It can be computer using the following formula [17]

$$\text{Coverage} = 100 * (\text{Pre-fetch Hits} / (\text{Pre-fetch Hits} + \text{Cache Misses}))$$

Pre-fetch can results in either of the following situation s

1. Useful Pre-fetch
2. Useless Pre-fetch

3. Harmful Pre-fetch

### 3.2.1. Useful pre-fetch:

Pre-fetch is useful when cache hit is the result before being supplant. Useful pre-fetch results in avoiding a cache miss

### 3.2.2. Useless pre-fetch:

Useless pre-fetch is the one which is when an item is supplanted before being accessed (pre-fetch miss). Its disadvantage is that it increases demand for cache bandwidth.

### 3.2.3. Harmful pre-fetch:

Unsafe pre-fetch is circumstance in which item is supplanted before being gotten to and pre-fetch suppleness a line that is solicited later (cache pollution). It results in an additional cache miss

Pre-fetch guidelines are executed by the processor to move data into the cache. The configuration/format of these guidelines looks like a typical load instruction, yet without a register specifier (since the information is just set in the cache). Pre-fetch guidelines likewise contrast from typical load guidelines in that they are non-blocking and they do not take memory special case “exceptions”. The non-blocking angle grants them to be covered with calculation, and the way that they do not take special cases is helpful in light of the fact that it allows more theoretical prefetching strategies (e.g., dereferencing pointers before it is sure that they indicate to legal locations).

The difficulties of programming controlled prefetching incorporate the way that some advancement is expected to embed the pre-fetches within the code, furthermore that the new pre-fetch instructions will include some measure of execution overhead. The benefits of programming controlled prefetching are that only a little amount of hardware backing is vital, and an extensive class of reference examples can be covered than simply steady stride accesses (e.g., circuitous references, for example in sparse-matrix code).

A pre-fetch algorithm should be precisely composed/designed if the machine execution is to be enhanced but not corrupted. To demonstrate this all the more plainly, we should first characterize our terms. Let the pre-fetch proportion be the proportion of the quantity of lines exchanged due to pre-fetches to the aggregate number of program memory references. And suppose transfer ratio be the total of the pre-fetch and miss proportions. There are two sorts of references to the cache,

1. Actual
2. Pre-fetch lookup

Genuine references are the ones that are produced by a source outer to the cache, for example whatever is left of the CPU (E-unit, I-unit) or the channel. A pre-fetch lookup happens when the cache checks itself to check whether a given line is available in it or in the event that it must be pre-fetched. The proportion of the aggregate accesses to the cache (real in addition to pre-fetch lookup) to the aggregate number of genuine reference is known as the access proportion/ ratio. There are distinctive sorts of costs connected with each of these proportions. These expenses can be characterized in as far as lost machine cycles per memory reference [18].

A cache pre-fetch algorithm contains 3 vital concerns:

1. At the point when to start a pre-fetch,
2. Line to pre-fetch
3. What substitution value to give the pre-fetched item/block

The proposed pre-fetching technique by us in which we used Probabilistic Graphical Model

(PGM). We have shown how we can prognosticate future items from the current ones and use this technique in cache.

Our proposed approach for pre-fetching is PGM, which gives us conditional dependency between random variables using directed or un-directed graph. PGM consist of two types "Markov chain" and "Bayesian network". **Markov chain** [19] says that the probability of next circumstance depends just on the present circumstance and not on the request of occasions that went before it. Similarly, **Bayesian network** speaks to an arrangement of arbitrary variables and show conditional dependencies between them using a directed acyclic graph. In our exploration, we exploited the Bayesian network technique to prognosticate the future solicitation by processor. We have shown results at the end which give us around 7% more cache hits then those traditional algorithms. The data-set which we used is **SMPCache** which is available open source mostly targeted for university level projects.

### 3.2.4. Pre-fetching Technique:

In this section, we will describe the methodology using which we implemented PGM. PGM is a way of representing probability distribution in a way that distribution can be represented such that we can read of conditional independent structure from a graph. They enable graph algorithms for inference and learning.

This is all about how we abstract the idea of conditional independence which is very statistical idea which is probabilistic and abstract in a way so that we can use computer science to think about how to efficiently reason about probabilities.

Example 1:

Below is a Bayesian network that describe conditional independent structure.

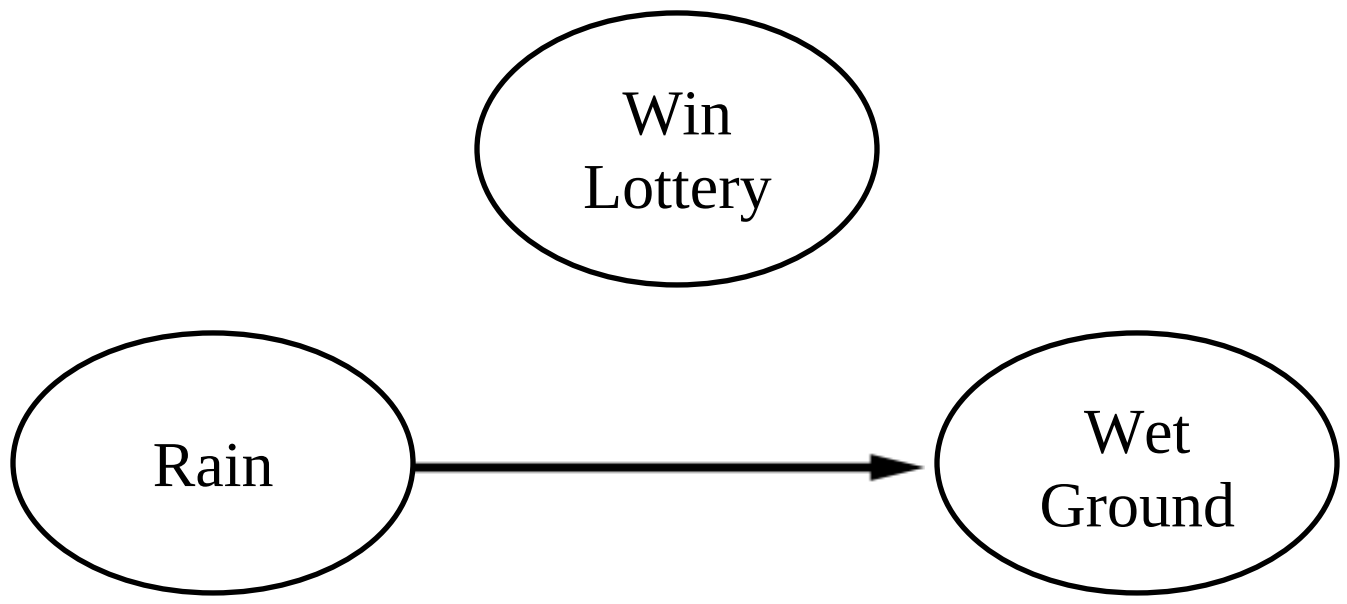

Figure 3.1 Bayesian network example

The distribution is represented by bunch of variables which are nodes and graphs, and directed edges between variables so that the directionality describes the conditional dependency.

P (L, R, W) = P (L) P (R) P (W|R)

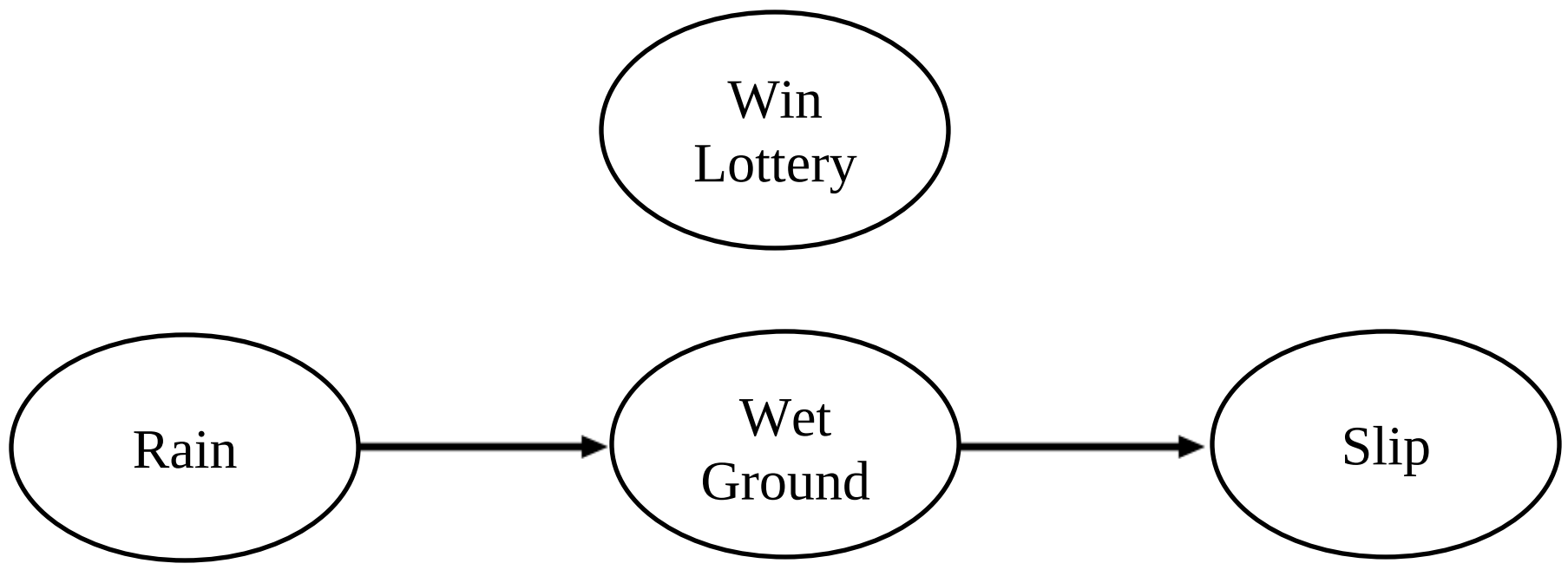

Figure 3.2 Bayesian probability

P (L, R, W, S) = P (L) P (R) P (W|R) P (S|W)

Bayesian network describes probability among all variables by putting edges between variable nodes.

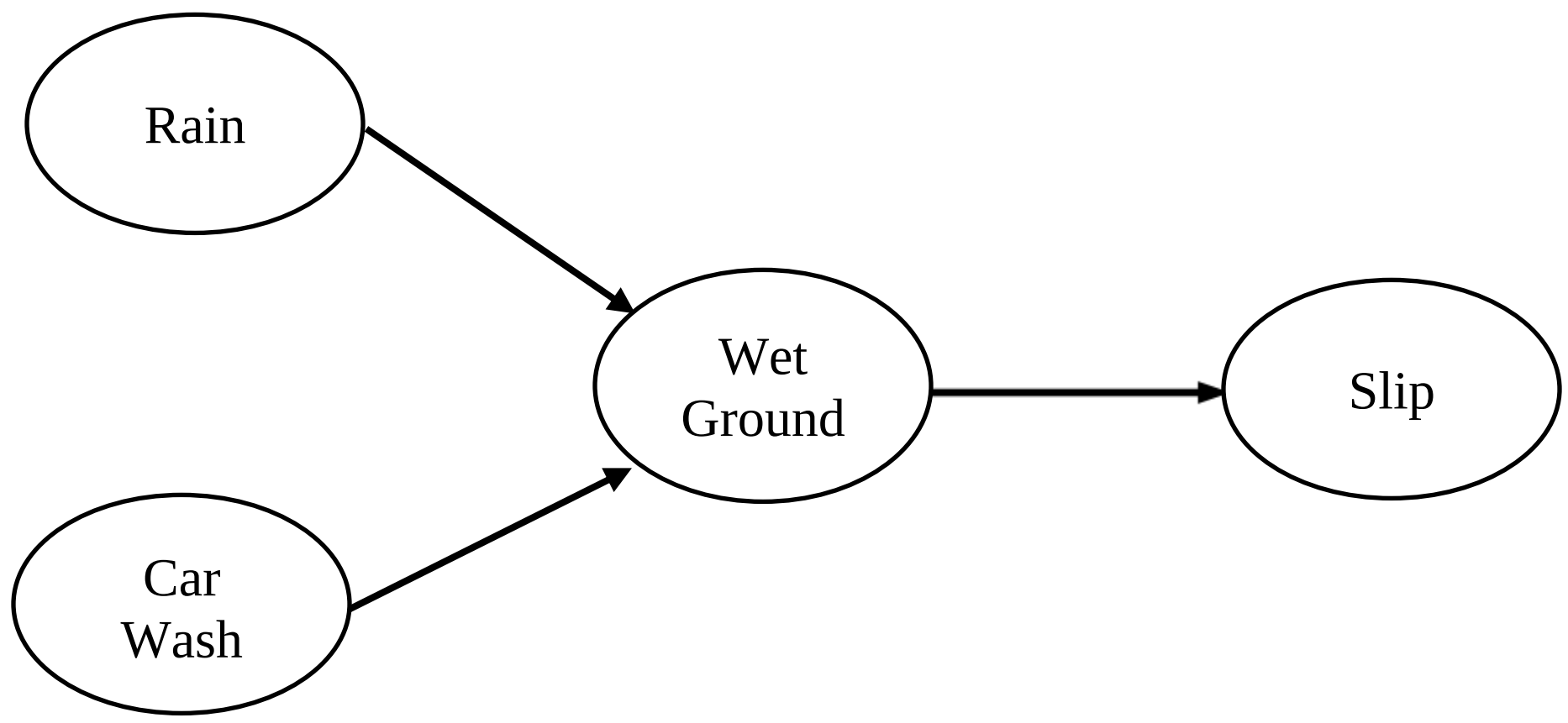

Figure 3.3 Bayesian extended probability

P (R, C, W, S) = P (R) P (C) P (W|R, C) P (S|W)

P[X | Parent(x)] => in above diagram, x is “Wet ground” and Parent(x) is “Rain” and “Car wash”.

Independence in Bayesian network says that each variable is conditionally independent of its non-descendent given its parent.

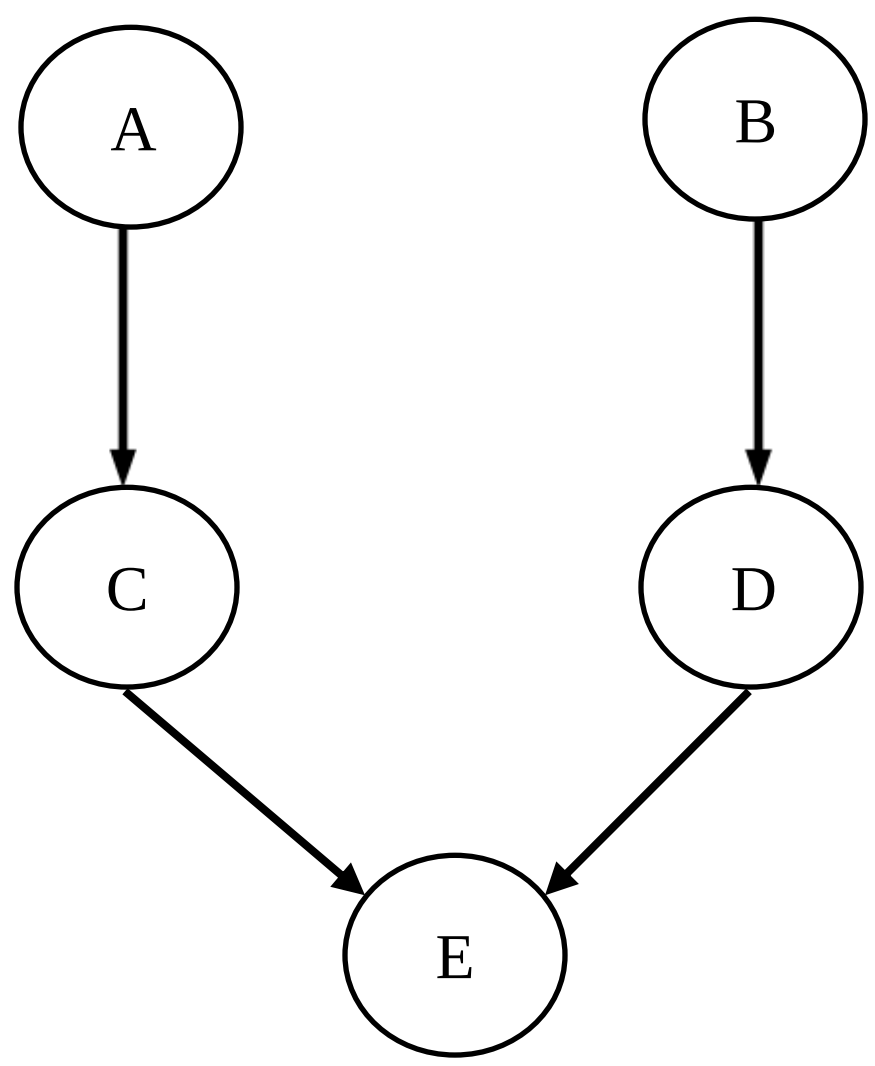

Figure 3.4 Independence in bayesian network

Now taking example of variable “C”, C has parent “A”. Now which variables are independent of “C”?

- "E” depend on “C”
- “B” and “D” are independent of “C”

Now taking variable “E”. “E” depends on “C” and “D” but it is independent of “A” and “B”.

Each variable is conditionally independent of any other variable giving its Markov blanket. In this case, node “C” has Markov blanket of nodes “A”, “E” and “D”. If we observed all the members of Markov blanket, then that variable become completely independent of any other variable in the network. That makes the things lot easier and lot more efficient to reason about.

There are two techniques which we considered during the process of our research

1. Enumeration
2. Variable elimination

## Inference:

Given a Bayesian network distribution P (X, Y, Z),

What is P (Y)? We will find out the answer to this question using enumeration and variable elimination.

### a) First Approach: Enumeration:

Sum over all possible settings of x and z, and all the remaining variables.

Taking the previously used diagram example.

$$P(R, W, S, C) = P(R)\,P(C)\,P(W|C, R)\,P(S|W)$$

Now let suppose we have to find probability of raining given that we slipped.

$$P(R \mid S)$$

Firstly we will use joint probability of all variables for all possible settings of “W” and “C”, because those are the variables we do not care about.

Then we have to divide the probability of “S”.

$P(R \mid S) = \sum_w \sum_c P(R, W, S, C) / P(S)$

$P(R \mid S) \propto \sum_w \sum_c P(R)\, P(C)\, P(W \mid C, R)\, P(S \mid W)$

This approach is expensive because we have to enumerate over all the possible settings of “W” and “C” and that is already in the example. Its cost is O (n). Now we have to consider how we can use Bayesian network structure to elevate this problem.

Now let us try to simplify this case. We have sum of “W” and sum of “C”. Now we will try to pull out terms that are not affected by variables being summed over. Now for example P(R) is nor effected by either “W” or “C”, so we will pull it all the way to the top.

$P(R \mid S) \propto P(R) \sum_w P(S \mid W) \sum_c P(C)\, P(W \mid C, R)$

Similarly, P (S | W) does not effected by “C”, therefore it is pulled out of “C” submission. And that is all for now as far as we can get.

Now we are down to slightly simplified form but still it is really expensive. Because in the formula above, P (W | C, R) has cost of $O(2^n)$.

We eventually going to sum this distribution P (W | C, R) which is a case where we have to look up at probability distribution over two variables and in general if we have “n” binary variables in the condition, we will need $2^n$ different probability distribution that we need to be look up, so that expensive. Now we have to try to avoid this.

Now if we view the given formula as tree with all possible settings of “W” and “C”.

$P(R \mid S) \propto P(R) \sum_w P(S \mid W) \sum_c P(C)\, P(W \mid C, R)$

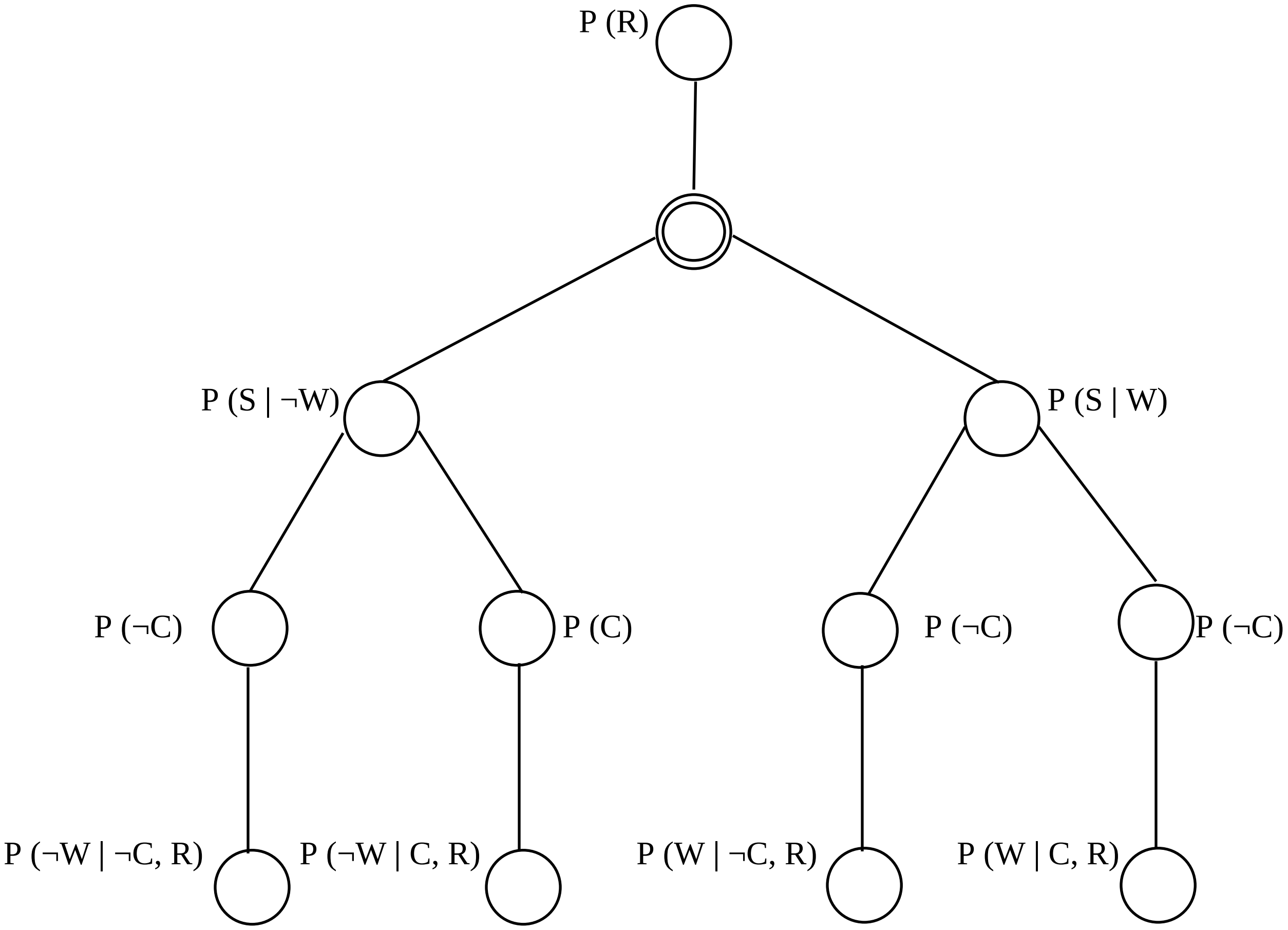

Figure 3.5 Enumeration

This is the idea of enumeration as shown in Figure 3.5. But generally it is wrong idea because it is not fully taking advantage of independence here. But it is useful to think about because it's the goal we are trying to accomplish, but we should do it in more efficient way.

b) Second Approach: Variable elimination:

At beginning, it looks almost same as enumeration. The formula according to Bayesian network is

$P(R \mid S) \propto \sum_w \sum_c P(R)\, P(C)\, P(W \mid C, R)\, P(S \mid W)$

Because of the way that Bayesian network has split up all those independent terms or conditionally independent terms, we can try to isolate one variable at a time.

Now for example let us look at "C" variable first. The terms depends on "C" variables are

$$P(C)\, P(W \mid C, R)$$

Now make function "C" that will take in same "W" which is the other variable it is depend on and output the sum-mission (Σ) over "C" for all those values.

$$F_C(W) = \sum_c P(C)\, P(W \mid C, R)$$

Now let us think of what the size of this function is. This function has to be computed by summing over all possible "C", so there is some cost here which is K in general.

Now we will write same formula again containing all the terms that did not depend on "C" and just add $F_C$ function with it.

$$P(R \mid S) \propto \sum_w P(R)\, P(S \mid W)\, F_C(W)$$

We replaced sum-mission (Σ) over "C" with function $F_C$.

## Variable Elimination Example:

We have probability distribution over four variables "W", "X", "Y", and "Z"

Now what if we try to find the probability of P(Y). The brute force technique for writing probability of P (Y) is

$$P(Y) = \sum_w \sum_x \sum_z P(W)\, P(X \mid W)\, P(Y \mid W)\, P(Z \mid Y)$$

Now let us run variable elimination method.

The idea is in start, we are going to pitch a variable, and then that variable will be eliminated by replacing that variable with the function.

$$P(Y) = \sum_w \sum_x \sum_z P(W)\, P(X \mid W)\, P(Y \mid Z)\, P(Z \mid Y)$$

Now first taking variable "W". We have to find out what are the terms in the joint probability distribution that depends on variable “W”.

$$F_W(X) = \sum_w P(W)\, P(X \mid W)$$

Now the new equation becomes

$$P(Y) = \sum_x \sum_z F_W(X)\, P(Y \mid X)\, P(Z \mid Y)$$

Now we eliminated “W” variable from above joint probability distribution function.

Now eliminating variable “X”

$$F_X(Y) = \sum_X F_W(X)\, P(Y \mid X)$$

In $F_X(Y)$, we used “Y” variable because variable other than “X”, which is used in the equation is “Y”, i.e. “Y” depends on “X”.

Now the equation becomes

$$P(Y) = \sum_Z F_W(Y)\, P(Z \mid Y)$$

Every variable that is not an ancestor of a query variable or evidence variable is irrelevant to the query. The process to iterate for variable elimination are

a) Choose variable to eliminate

b) Sum the terms relevant to variables, generate new factors.

c) Perform elimination until no more variables to eliminate

Now the question is that how do we do learning in Bayesian network. Assume that we fully observed situation, we have bunch of training examples that tell us the state of every single variable in our distribution. That is the case where learning becomes super easy. And we already saw this in naive Bayes. It was all counting. Estimate each conditional probability like we did in naive Bayes. All we do is just count the probability of each variable given its parents from our training set. And then we might add same priority like we did for naïve Bayes to prevent over fitting or to prevent situation where we do not have enough data.

# Chapter 4 Experiments, Results and Discussions

The study which has been done in this chapter includes the results of experiments conducted on **SMPCache** dataset. We conducted multiple experiments in which we firstly implemented classical algorithms (FIFO, LIFO, LRU, MRU), then we implemented our own algorithm (PMG), then we have shown difference between our algorithm and those classical ones by comparing the results of both. At the end, we have shown how our approach is better than all those.

## 4.1. Data set:

In our experimental setup, we varied the size of cache (k) to be “n”, “log (n)”, “Under root (n)”. The data-set used in the setup is **SMPCache** which is available open source. This dataset is utilized by an author whose name is William Stallings. The dataset has been used as simulation tool for usage of understudy ventures in a book named as “Computer Organization and architecture”, Prentice-Hall, sixth edition, 2003. The authors of this dataset are Miguel A. Vega-Rodriguez [20]

## 4.2. Experimental Setup:

In Experimental setup, we selected few of the several techniques which are effective in cache field of domain. Then we tried to analyze those techniques by performing different types of experiments. Experiments are carried out by giving then varied dataset and cache size and then observed their behavior and possible reason of that behavior. Then we tried to analyze those behaviors and tried to overcome those limitations in our customized algorithm. In this way only, we became able to design our algorithm which can gives us desired results. The classical algorithms which are selected for our experiments was (LIFO, FIFO, LRU, and MRU). The reason for selecting those algorithms is that according to our knowledge, these were the techniques which was using mostly in modern research. This has helped us to evaluate modern trends in cache memory research.

We used probabilistic graphical modeling algorithm to perform our experiments. The language used is Java. Compiler used is Eclipse/Netbeans.

## 4.3. Results and Discussion:

In the coming section, the results of our experiments have been shown and also discussed the behavior of algorithms in different scenarios and their impact.

### 4.3.1. Scenario 1:

In first scenario, we have taken k = log (n), which is k=6 when size of data taken for this experiment is 600,000

By performing this experiment, we tried to analyze the behavior of classical algorithms that how they perform when we vary the size of dataset. Every algorithm from those given bellow have some limitations which are shown in experimental results. As we can see from the results, LRU contain most hits from all those. But if we look overall results, then there are a lot of cache miss which is a bit of a worry factor because it will gradually effect the performance of processor.

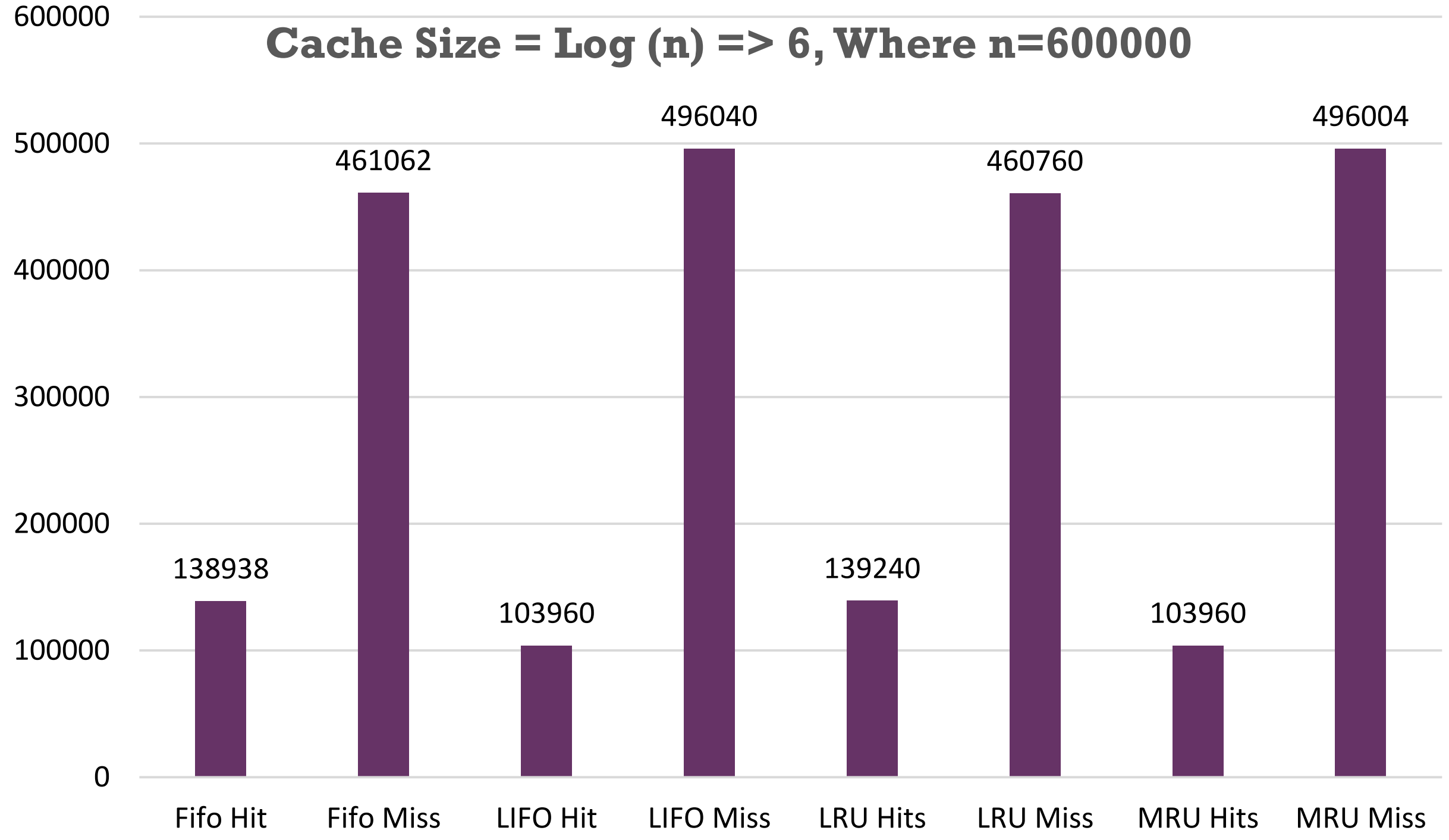

Figure 4.1. Classical algorithm example

The results shown in Figure 4.1 are taken after experimenting those classical algorithms by giving them our dataset. We can see how much hits and miss every technique achieves. On the basis of those results, we can easily predict which algorithm is best amongst them. Then we evaluated that how that best technique among them is unique from other techniques. This helped us to design our own algorithm. The reason of very few hits and large number of miss is the small size of cache. This behavior clearly tells us that size of cache effect the total number of hits given by any technique. The size of cache is small in start and increased as we go further forward in order to observe its behavior with different sizes.

### 4.3.2. Scenario 2:

In second scenario, we have taken k = Under Root (n), which is k=775 when the size of data taken for this experiment is 600,000.

In this experiment, if we compare it with scenario one, we can see that there is huge amount of increase in cache hit. The main reason for this behavior is the increase in the size of k. If we increase the size of k, it give us additional space to store more items in the cache at same time, in this way, there is high probability for processor to find the item in cache regardless of the fact that which technique is used to arrange data in cache memory. But in reality, it is really difficult to increase the size of cache memory because of high cost factor. Therefore, our basic purpose of this research is to retain the size of cache as minimum as we can and still try to get maximum hits, and also keeping in mind the overhead factor. Another thing which is interesting about this scenario is that again LRU is giving more hits as compare to other algorithms. The reason of this behavior is that LRU remove the item from cache which is called least number of time from all items which are present in cache at that time. Therefore, it is an optimum technique from all those which we considered in our experiment.

In Figure 4.2, everything is same as in scenario 1 except the size of cache. In this case, size of cache which we have taken for our experiment is $\sqrt{n}$. This makes the cache so large and because of that, we observed such huge amount of increase in cache Hits. This figure support our point which we made in start that cache hits increase as we increase the cache size.

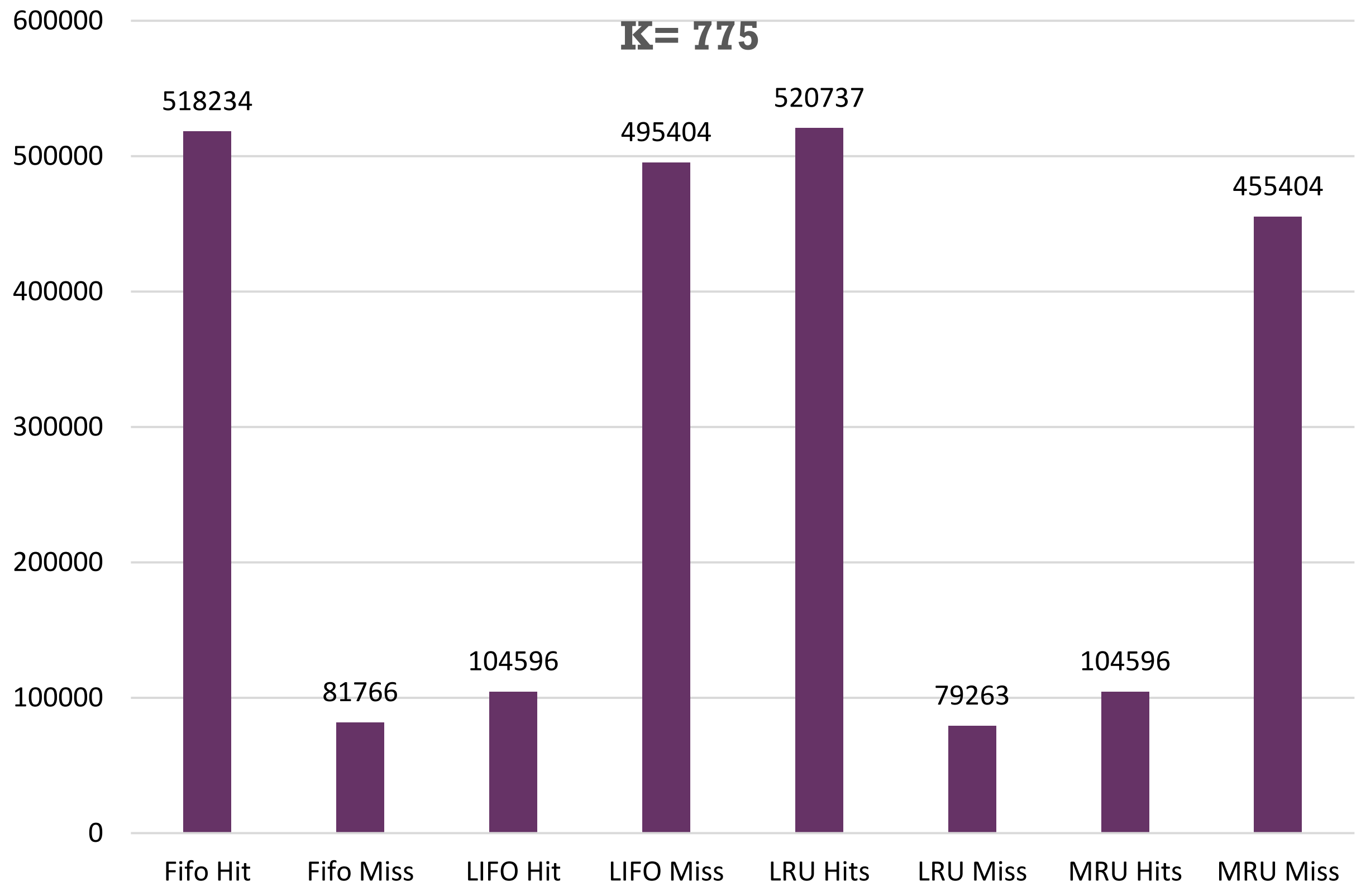

Figure 4.2. Implementing classical algorithms with k = √ n

### 4.3.3. Scenario 3:

In second scenario, we have taken k = 32 when the size of data taken for this experiment is 600,000.

We varied the size of cache which not changing the size of dataset. The reason for doing this is to analyze the behavior of classical algorithms and their impact on different cache sizes. Again the common thing in this scenario from the above scenarios is the most number of Hits by LRU as compared with other algorithms. But we can clearly observe the decrease in the overall cache hits of all algorithms. This behavior shows that cache with less size will have most cache miss if we do not organize the items in proper order.

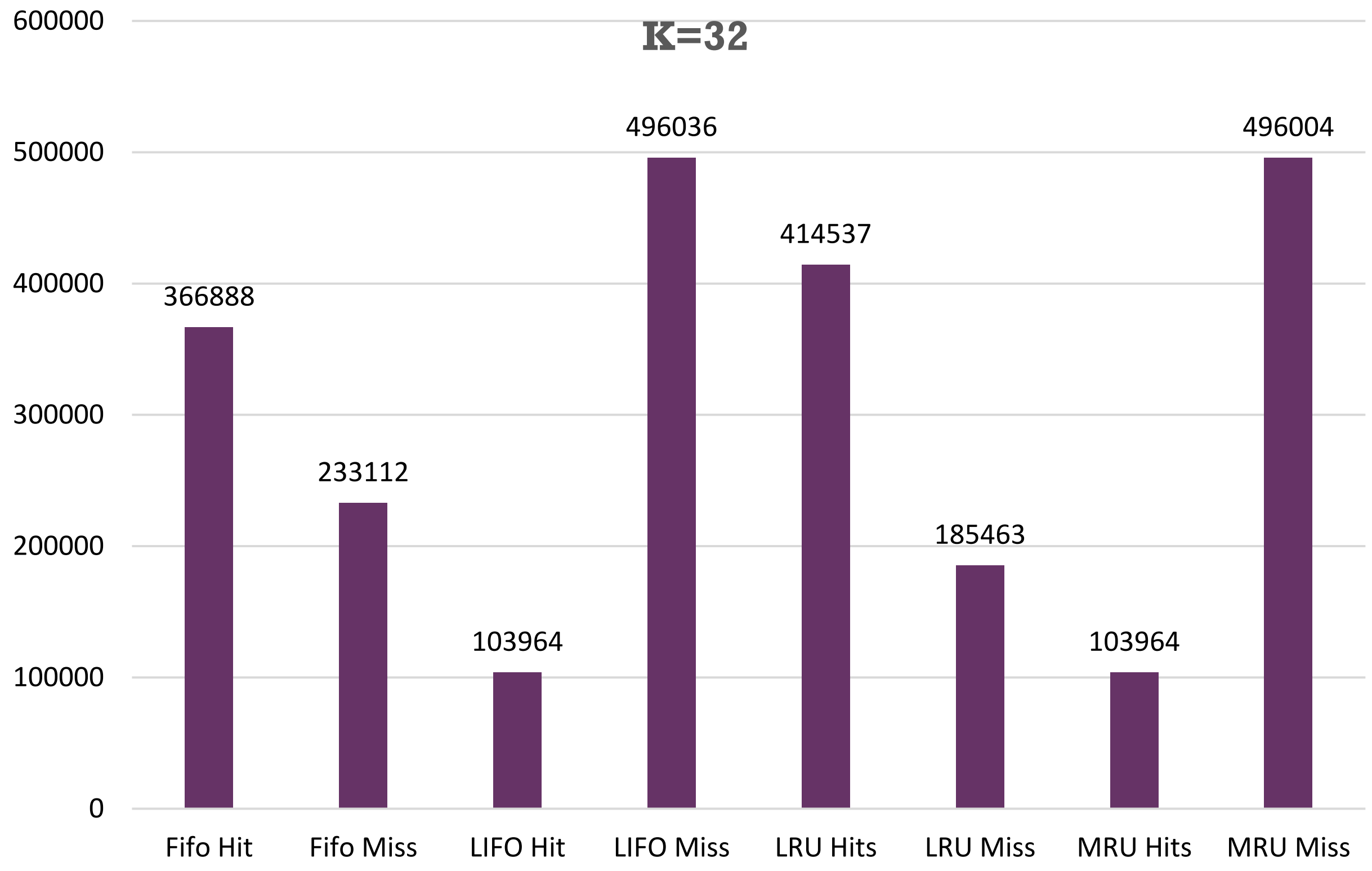

Figure 4.3. Implementing classical algorithms with k = 32

Figure 4.3 contains same experimental setup as in scenario 1 and scenario 2. The only difference is in size of cache which gives us the following results which are shown in figure. Again by reducing the size of cache, we observed a decline in total cache hits. Another thing to be consider is that LRU technique has given best results in all three scenarios. This proves it to be better technique then others.

### 4.3.4. Scenario 4:

In this scenario, we experimented our dataset with our customized algorithm which is based on PGM. The size of cache in this experiment is k=3, and size of dataset is 1000

The reason for taking small dataset was to check in start that whether our technique is effective on small scale or not. Firstly we experimented that dataset on our algorithm without using PGM technique by placing values randomly and noting the results. Then we experimented same data by applying the PGM technique. We can clearly see the difference as we were getting 7% increase in cache hits. This experiment shows that there is potential of using PGM for our experiment and that encourages us to research further in this field.

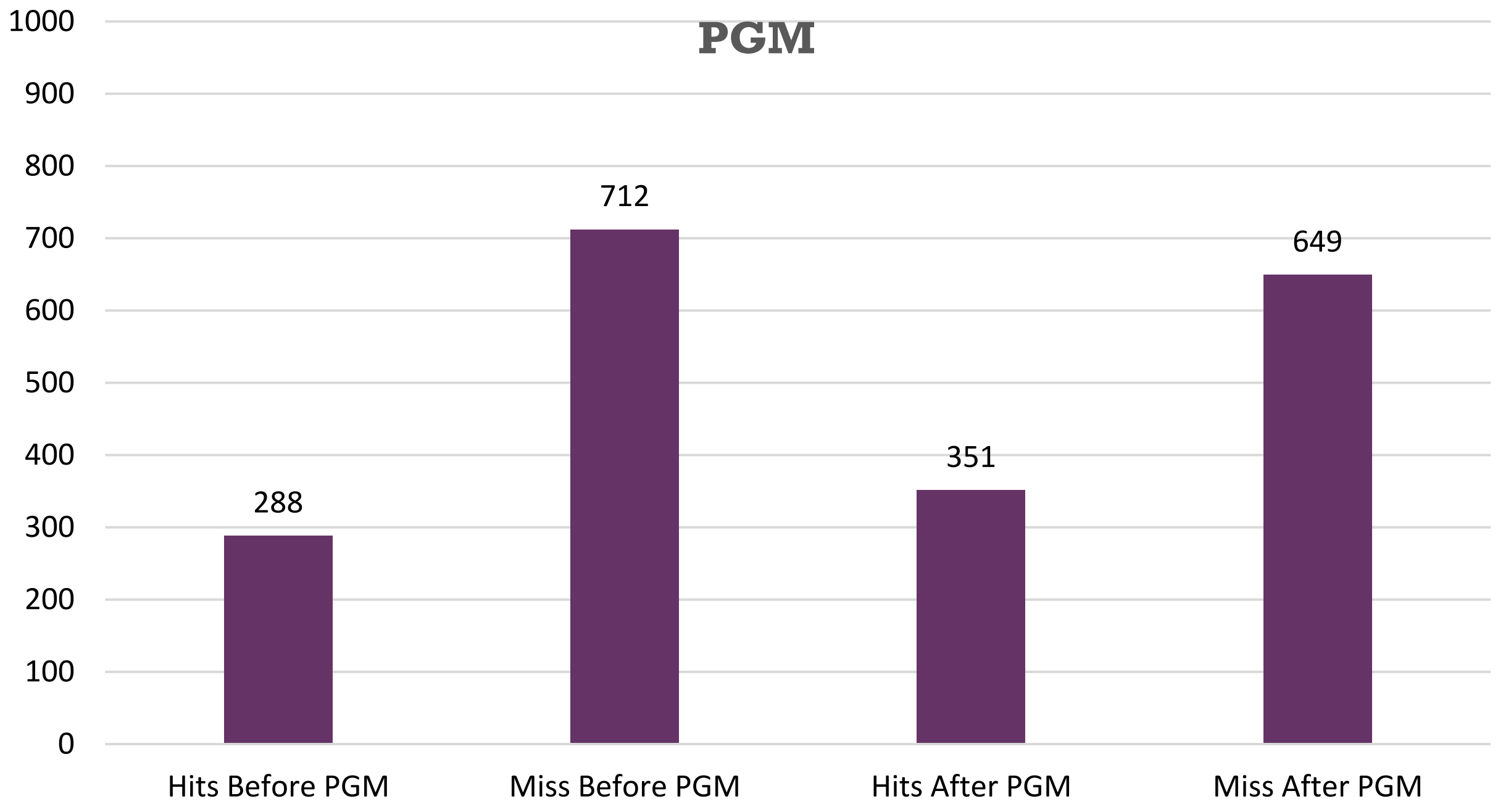

Figure 4.4. Implementing customized algorithm with and without PGM

Figure 4.4 contains the results of our customized algorithm firstly without using PGM technique and then after using PGM technique. The dataset in this figure is really small (1000 traces). The reason was to first verify that is PGM technique is capable enough to solve our needs or not. This question is clear after seeing this figure that PGM contains high potential to be one of the future technique for cache algorithms.

### 4.3.5. Scenario 5:

In this scenario, we experimented our dataset firstly on classical algorithm without using PGM technique, then with PGM technique, and then we compared the results of those behaviors. We can see from the results that PGM gives more hit on those classical techniques if we apply

PGM on them. The size of dataset is in this experiment is 80,000.

These results show that our technique is better than those classical ones.

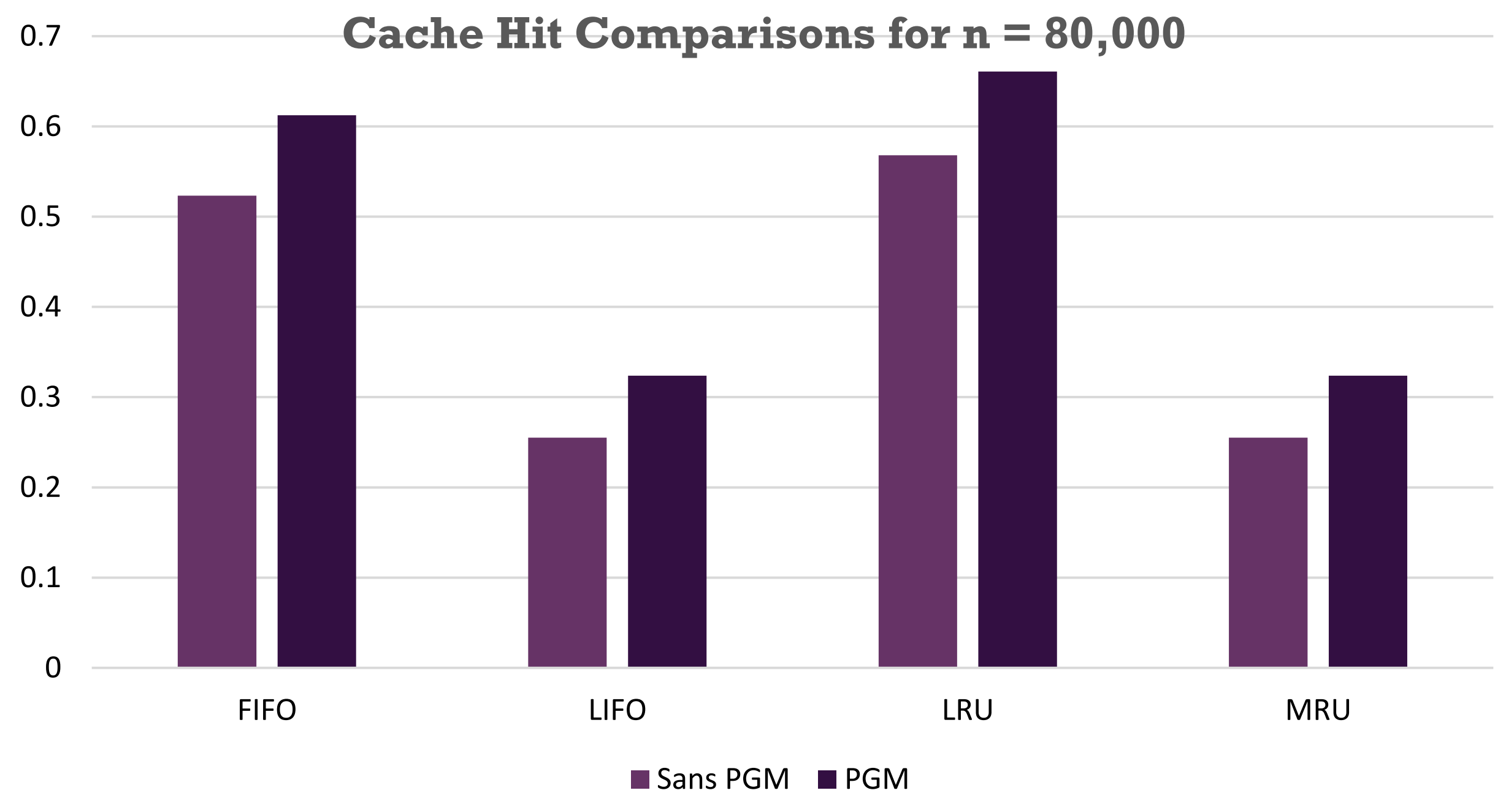

Figure 4.5. Comparing classical algorithms hits with and without PGM

Figure 4.5 contains the mixture of classical algorithms and PGM technique. This figure clearly verifies our claim that PGM technique is better that those classical ones. The hits after applying PGM technique to those classical ones has increased which can be seen clearly in the figure.

## 4.4. Conclusion:

From those results shown above, there are a lot of things to be considered. We analyzed the behavior of different types of algorithms. It has been shown that different algorithms perform better in different scenarios, also they failed in certain scenarios. Our goal was to minimize the possible risk of failure in those certain scenarios which maintain the best performance in all those ideas scenarios.

The first thing to be considered from those experiments is that if we increase the size of cache, then the hit rate increases gradually. But the problem is the cost factor due to which we cannot increase the size of cache. Therefore, the possibility of getting high hit rate by increasing the size of cache was neglected.

Second thing which is observed is the high hit rate of LRU then other techniques. “A decent approximation to an ideal algorithm depends on the perception that things that are vigorously utilized as a part of the last couple of instructions will likely be intensely utilized again as a part of the following few. On the other hand, things that are not utilized for very long time will most likely stay unused for quite a while. This thought proposes a feasible algorithm: when a thing flaw “item fault” happens, erase the item that has been unused for the very long time. Hence this philosophy is known as **LRU** (**Least Recently Used**) technique [20]”.

If we take example of LRU, it takes the cache block which is least used and suppleness it with the item which has the highest probability to be called next. Generally this technique make sense but what is that the block which is used least can be called next by processor? In that case, we will have to face a cache miss which will cause additional time to fetch that item from main memory. Therefore, we can say that from those all classical algorithms used in our experiments, LRU suppleness the item more intelligently then others.

Third thing which is observed is that PGM gives more hits than others. This technique prognosticate future items by using their conditional dependencies. We exploited the terminology “Markov blanket” [21] which says that the probability of the next item solely depends on its “Parents”, “Child”, and “Child's parents” only. This technique pre-fetch the items from main memory before processor calls for those items. In this way, processor finds that item in cache most of the time without going into main memory which at the end increases hit rate.

# Chapter 5 Conclusion and Future Work

This thesis shows that by using different technique other than those traditional ones, we can obtain desire results by using them intelligently.

We proposed two techniques:

1: Pre-Eviction

2: Pre-Fetching

Further there is a lot margin of improvement available in cache field which needs to be exploited as shown in our research that arranging items intelligently can gives us desire results. Things which ought to be kept in mind is that we have to keep in mind those scenarios in which there is the possibility of failure of our algorithm and we have to work on those scenarios to minimize their effect as possible. In this way we can increase the hit rate and overall performance of the processor by keeping it busy all the time.

Also there is not important by trying to optimize the algorithm by using one technique. We can combine multiple techniques (Hybrid) to get our desire results if that cannot cause high overhead.

## 5.1. Future Work:

Further in future we can extend our research to perform further experiments to analyze the behavior of cache so that it can be optimize further to produce better results. If we design a hybrid approach which uses PGM along with other techniques intelligently then we can increase the cache hits which will results in the fast processor performance.

While cache hit ratios are ordinarily used to gauge the execution of a cache algorithm, we will likewise contemplate in future the total measure of time which has been spent waiting for record access occasions to finish. The main purpose is to extend the simulation to incorporate read- hold up time which allow us to represent for the extra Input/output load created by prefetching.

At long last we mean to investigate a prefetching determination/selection model which uses the estimated probabilities of model given by us, in conjunction with different multiple factors, for example memory pressure and document or file size, to appraise the expense, in terms of read- hold up time. These estimates would be utilized as a part of every case to choose in the event that it was more gainful to pre-fetch or not.